\documentstyle[preprint,aps]{revtex} 

\begin{document}



\nopagebreak
\title
{
Ising Model formulation of Large Scale Dynamics:\\
Universality in the Universe. \\
}

\author{J.\ P\'erez--Mercader$^{*}$, T. Goldman$^{\dagger}$,  D.
Hochberg$^{*,\dagger\dagger}$ and R. Laflamme$^{ \dagger}$
\\
\vspace{.125in}}

\address{
$^{*}$Laboratorio de Astrof\'{\i}sica Espacial y F\'{\i}sica
Fundamental\\
Apartado 50727\\
28080 Madrid, Spain\\
$^{\dagger}$Theoretical Division\\
Los Alamos National Laboratory\\
Los Alamos, New Mexico 87545, USA\\
$\dagger\dagger$Departamento de F\`{\i}sica Te\'orica\\
Universidad Aut\'onoma de Madrid\\
Cantoblanco, 28049 Madrid, Spain\\
}

\maketitle

\vspace{-13.5cm}
\begin{flushright}
LAEFF--95/004\\
LA--UR--95--1055
\end{flushright}
\vspace{12cm}

\vspace{-.5in}
\begin{abstract}

The partition function of a system of galaxies in gravitational
interaction can
 be cast in an Ising Model form, and this reformulated via a
Hubbard--Stratonovich transformation into a three dimensional
stochastic and
classical scalar field theory, whose critical exponents are
calculable and
known. This allows
one to $compute$ the galaxy to galaxy correlation function, whose
non--integer
exponent is predicted to be between 1.530 and 1.862, to be compared
with the
phenomenological value of 1.6 to 1.8.
\\
\end{abstract}



{\sl PACS: Self--Gravitating Systems: 04.40; Cosmology: 98.80;
Renormalization,
-- phase transitions: 64.60A.}

\vspace{.5in}


An intergalactic tourist admiring the Universe at the
largest scales would perceive it as something akin to a three dimensional
`salt--and--pepper' pattern which, when projected onto a two dimensional
picture, would appear very similar to what $we$ see in pictures of
Large Scale
Surveys, such as the Lick$^{\cite{lick}}$ or APM surveys$^{\cite{apm}}$.
As he reduced the size of his gauge to smaller and smaller distances
he would come to the conclusion that the Universe today is dominated by
matter which, at large distances is in gravitational interaction.
And that this
matter seems to organize itself into bodies which roughly group
themselves into
solar systems, galaxies, groups of galaxies, and even larger
structures. We
$believe$ we see the same as our tourist, and that we call the
Cosmological
Principle.

At the larger scales he would probably recognize  the Universe as a
homogeneous
object and therefore describe it with a
Friedmann--Robertson--Walker metric. He
would reproduce the many successes of cosmology based on this
metric. As he
went on to smaller scales he would find that (i) {\it `on {\it all}
observable
scales there are structures seen and significant anisotropies are
detected'}$^{\cite{ostriker}}$. In fact, (ii) if he inferred the
galaxy--to--galaxy
correlation function ($\xi_{Gal}(r)$) from these surveys, he would
discover\footnote{The value of $\gamma$ seems to vary somewhat from
survey to
survey; for example, it is 1.8 for the Lick survey and about 1.6
for the APM.}
that $\xi(r)
\propto r^{-\gamma}$ where$^{\cite{peebles}}$ $\gamma \sim O(1.6 - 1.8)$,
instead of $\gamma =1$
which is what one would naively expect\footnote{This can be understood as
follows: as we will see below, the gas of galaxies can be put in a
one--to--one
correspondence with a 3--dimensional scalar field theory, and the
galaxy--to--galaxy correlation function corresponds to the scalar field
propagator. In units of length, a scalar field in a $d$--dimensional
space--time has a canonical dimension of $-(d/2 - 1)$. The above
statement
about $\gamma$ follows at once.} for a homogeneous distribution of
matter in a
three dimensional $space$.

Given these phenomenological facts, one may ask: (A) Is
there a $fundamental$ explanation for this power law behavior? (B)
Can it be
understood by using some basic $scheme$?

This note aims at providing answers to the above questions within
the framework
of known physics. It rests upon the well--known
observation$^{\cite{wilson}}$ that the existence of a non--integer
(anomalous)
dimension (signalled by $\gamma \neq 1$) is a tell--tale sign
betraying the
existence of $both$ smaller length scales and fluctuations.

We will apply the techniques of  Statistical Mechanics to a system
made up by
many--galaxies (accounting for the `smaller length scales' )  in
gravitational
interaction, and which are subject to fluctuations  emerging from
the intrinsic
properties of the gravitational interaction of this many body system. The
actual nature of the fluctuations needs not be $specified$ here, but they
could, for example, be related to the `frictional' processes in
gravitational
systems long ago considered by Chandrasekhar$^{\cite{chandra}}$ or, as
predicted by well known classical theorems of Poincar\'e, to chaotic
processes$^{\cite{henon}}$ related to the many body nature of the
gravitational
system.

We will obtain the partition function for this system and compute the
two--point correlation function and its corrections due to the
existence of
fluctuations.

Let us consider the continuous mass density, $\rho(\bf{r})$,
describing the
spatial--distribution of galaxies. The deviation from an average density
$\bar{\rho}$ is $\delta \rho(\bf{r})=\rho(\bf{r})-\bar{\rho}$.  The
galaxy--to--galaxy correlation function is defined$^{\cite{peebles}}$ as
$\xi_{Gal} ({\bf{r}}_i-{\bf{r}}_j) = \left\langle  \delta
\rho({\bf{r}}_i)
\delta \rho({\bf{r}}_j)\right\rangle / \bar{\rho}^2$, where the angular
brackets mean that a suitable average has been taken. In the
(justifiable)
non--relativistic limit, the gravitational
interaction energy for this system is then given by (assuming no
expansion)

$$
H_{int}={ -1\over 2}\,G\,\int\int
d^3{\bf{r}}_i
d^3{\bf{r}}_j
\rho({\bf{r}}_i)\frac{1}{|{\bf{r}}_i-{\bf{r}}_j|}\rho({\bf{r}}_j)\,.
$$

As a first approximation, it is reasonable to consider the gas of
galaxies as
made up of discrete, spatially localized `points' of (equal) mass
$m_0$, and we
can set$^{\cite{peebles}}$
the `contrast'  $\delta \rho({\bf{r}}_i) / \bar{\rho}$ equal to 1
if there is
a galaxy at position ${\bf{r}}_i$ and equal to --1 if there is a void.
The interaction energy becomes

\begin{equation}
H_{int}=-\frac{1}{2} \sum_{ij} m_i\frac{G}{|{\bf{r}}_i-{\bf{r}}_j|}m_j
 \equiv -\frac{1}{2} \sum_{ij} m_iL_{ij}m_j
\label{1}
\end{equation}
with the following natural correspondence (since we have assumed that all
galaxies have a similar mass) between $m_i$, a two--valued $(\pm
1)$ `$spin$'
variable $s_i$, the density and the contrast:

\begin{equation}
2\,\,m_i/m_0 \leftrightarrow \rho({\bf{r}}_i)/\bar\rho
\label{2a}
\end{equation}

\begin{equation}
s_i  \leftrightarrow \delta \rho({\bf{r}}_i)/\bar\rho.
\label{2b}
\end{equation}

Furthermore, $m_i$ and $s_i$ are related by

\begin{equation}
m_i =m_0\,\frac{1}{2}(s_i+1)
\label{trans}
\end{equation}
which happens to be analogous to the relationship between a lattice
gas and an
Ising magnet.

This relationship, via a Hub\-bard--Strato\-no\-vich transformation
$^{\cite{block}}$,
allows one to map the `gas' of galaxies into a system described by
a stochastic
1--component (scalar) classical field $\phi ({\bf{r}})$ in 3
dimensions, whose
partition function $Z[\beta]$ can be readily calculated, and gives

$$
Z^{Grav}_{Ising}\left[ \beta \right]= \sum_{\{m\}}
{
e^{
\frac{\beta}{2}\sum_{ij}{m_iL_{ij}m_j}}
}
$$
$$
=C \,\int{[d\phi] \exp{\left\{-\frac{\beta}{2}\int{
\left(\phi(r)-H(r)\right)
L^{-1}(r,r^\prime)\left(\phi(r^\prime)-H(r^\prime)\right)
d{\bf{r}} d\bf{r^\prime}}+
\int d{\bf{r}} \log{\cosh{[\beta \phi(\bf{r})}}]\right\}}}
$$
\begin{equation}
\equiv \int{[d\phi] e^{-\beta {\cal H}[\phi,H]}} \, .
\label{3}
\end{equation}

Here the function $H(r)$ is given by $H(r)=-1/2 \int
d{\bf{r^\prime}} m_0^2\,
L(r,r^\prime)$ and $C$ is an inessential factor. The field $\phi$
is the order
parameter for this system.
Although all direct reference to the original masses has disappeared, the
physics described by the hamiltonian of Eq. (\ref{3}) is completely
equivalent
to the original description.

Because of Eq. (\ref{1}) and (\ref{trans}),

$$L^{-1}(r,r^{\prime})=
-[\delta (r-r^{\prime})/(2\pi Gm_0^2)] \nabla^2_r \, ,
$$
and

$$
 {\cal H}[\phi,H]=-\frac{1}{2} \int{
\left(\phi(r)-H(r)\right)
\frac{1}{2 \pi Gm_0^2} \nabla^2
\left(\phi(r)-H(r)\right)
d{\bf{r}}}
$$
$$
-\frac{1}{\beta}\int d{\bf{r}} \log{\cosh{[\beta \phi(\bf{r})]}} \, .
$$

As is well known$^{\cite{block}}$, the connected, two--point correlation
function for the spin system and the field theory are the same.
Furthermore,
because of fluctuations in the field $\phi$, its canonical
dimension acquires
an {\it anomalous dimension} and shifts away from its canonical
value (cf.
Footnote 2), such that when $|{\bf{r}}_i - {\bf{r}}_j| \rightarrow
\infty$, the
connected, 2--point
correlation function for $this$ hamiltonian, $\xi (|{\bf{r}}_i -
{\bf{r}}_j|)$,
scales as$^{\cite{wilson}}$$^{\cite{block}}$

\begin{equation}
\lim_{|{\bf{r}}_i - {\bf{r}}_j|\rightarrow \infty}
\left\langle s_i \, s_j \right\rangle =
\lim_{|{\bf{r}}_i - {\bf{r}}_j|\rightarrow \infty}
\xi (|{\bf{r}}_i - {\bf{r}}_j|) \sim {\frac{1}{|{\bf{r}}_i -
{\bf{r}}_j|^{d-2+\eta}}}
\label{5}
\end{equation}
where the first equality follows from the equivalence between the
`$spin$' and
`$field$' descriptions of the system, $d$ is the dimensionality of
space (=3)
and $\eta$ is the critical
exponent for the pair correlation function, whose value (0.0198 --
0.064)  (cf.
the Table below)
differs from zero due to the fluctuations in $\phi({\bf{r}})$.

Because of Eq. (\ref{2a}) and (\ref{2b})

\begin{equation}
\xi_{Gal} (|{\bf{r}}_i-{\bf{r}}_j|) =
\left\langle s_i \, s_j \right\rangle
\label{6}
\end{equation}
with the average computed using Eq. (\ref{3}).

Putting together Eqns. (\ref{5}) and (\ref{6}), our calculation
$shows$  that
for large separations, the galaxy--to--galaxy correlation function
{\it must}
scale  as

$$
\xi_{Gal} (|{\bf{r}}_i-{\bf{r}}_j|) \sim r^{-\gamma}
$$
with$^{\cite{barber}}$ $\gamma = d-2+\eta$ between 1.0198 and 1.064.

Thus far our calculation has been static, but the Universe is
expanding, and
the effects of expansion can (and do$^{\cite{goldenfeld}}$) modify
the values
of critical exponents. For the correlation function, it is known
from computer
simulations in condensed matter physics$^{\cite{goldenfeld}}$
combined with
dynamical scaling considerations, that time enters in the
correlation function
by altering the argument of the correlation function from $|{\bf{r}}_i -
{\bf{r}}_j|$ to  $|{\bf{r}}_i - {\bf{r}}_j|/L(t)$ where $L(t)
\propto t^\zeta$
and $\zeta$ is determined in computer simulations to be 1/3 for
systems with a
conserved order parameter, and  1/2 for systems with a
non--conserved order
parameter. Separation and time are related in an expanding Universe
where, to a
first approximation, in a {\it matter dominated Universe} the scale
factor is
proportional to $t^{2/3}$. Putting together the expansion of the
Universe and
the dynamical critical phenomena effects as contained in $\zeta$,
the exponent
in the galaxy--to--galaxy correlation function is modified from
$\gamma = d-2
+\eta$ to $(d-2+\eta)\times(1+3 \zeta/2)$. That is, we finally get
that the
predicted (calculated) value for $\gamma$ will be\footnote{Cf. Table. The
static values for $\eta$ are taken from Reference \cite{barber}.}
between 1.530
and 1.596 ($=\gamma_{Expanding}^{C}$) if we assume that the order
parameter is
conserved, and between 1.785 and 1.862 ($=\gamma_{Expanding}^{NC}$) if we
assume that the order parameter is not conserved.

\vspace{1.5cm}
\begin{tabular}{||c|c|c|c||}
\hline \hline
Method of Calculation & $\gamma_{Static}$ & $\gamma_{Expanding}^{C}$ &
$\gamma_{Expanding}^{NC}$\\
\hline \hline
Series Estimates &1.056 $\pm$ 0.008 & 1.584 $\pm$ 0.012 & 1.848
$\pm$ 0.014\\
O($\epsilon$)              &0                   &   1.5      & 1.75 \\
O($\epsilon^2$)         &1.0198         &   1.530 & 1.785 \\
O($\epsilon^3$)         &1.037           &   1.555 &  1.815 \\
O($\epsilon^4$)         &1.029           &   1.543 &  1.801 \\
\hline \hline
\end{tabular}
\vspace{1.5cm}

These values are to be compared with the values inferred from the
existing
galaxy catalogs, which range between 1.5 for the APM survey to 1.8
for the Lick
survey$^{\cite{peebles}}$.

Therefore, we see that the questions enumerated at the beginning of
this note
can find an answer within the framework outlined here. In addition,
there is a
clear and unambiguous prediction: due to the Universal nature of the
gravitational force, reflected in the interaction hamiltonian of
Eq. (\ref{1}),
the result  we have obtained for the galaxy--to--galaxy correlation
function
must apply also to {\it any other many--body--gravitational
system}, including
the interstellar medium in our galaxy. This means that observations must
confirm that the interstellar medium has a distribution whose correlation
function scales with the same $generic$ power law as galaxies. The only
possible difference would be in the numerical value of the
anomalous dimension,
since for intergalactic gas clouds the size of the system is smaller, and
therefore, renormalization group arguments tell us that the value of
$\gamma_{Interstellar}$ is smaller than for systems of galaxies,
where the
coupling constant has had more distance to grow on its way into the
IR fixed
point\footnote{This follows by noticing that, because of conservation of
probability (unitarity in field theory) $\eta$ is positive;
perturbation theory
tells us that it is  proportional to a power of $Gm_0^2$. Also, in
less than
four dimensions, the latter coupling tends in the IR to the
equivalent of the
Wilson--Fisher fixed point, and away from the Gaussian UV--fixed point.}.

Many questions remain. For example, is the order parameter
conserved, or not?
Can one use renormalization group techniques to also $calculate$
the dynamical
effect of the expansion of the Universe on $\eta$, instead of
appealing to
phenomenological (albeit well substantiated) computer estimates to
generalize
the static values of $\gamma$ to dynamical values? Does the implied
$r$-dependence of $G(r)$ play a r\^ole in the physics of large
scales? How do
initial conditions impact on the correlation functions? How does
one actually
approach the disordered phase? These questions will be considered
in a future
paper.

\end{document}